\documentstyle[twocolumn,fleqn,psfig]{article}

  \advance\oddsidemargin by -1in
  \advance\oddsidemargin by -6mm
  \advance\evensidemargin by -1in
  \advance\evensidemargin by -6mm
  \advance\topmargin  by -1in
  \advance\topmargin  by 3mm
\def\thebibliography#1{\section*{References}
 \list
 {[\arabic{enumi}]}{\settowidth\labelwidth{[#1]}\leftmargin\labelwidth
 \advance\leftmargin\labelsep
\setlength{\itemsep}{0ex}  
\setlength{\parsep}{0ex}   
 \usecounter{enumi}}
 \def\newblock{\hskip .11em plus .33em minus .07em}
 \sloppy\clubpenalty4000\widowpenalty4000
 \sfcode`\.=1000\relax}

\begin{document}

\twocolumn[


\begin{center}

{\bf cond-mat/9807047; v.1 July 2, 1998; v.2 September 25, 1998}

  {\it To be published in Synthetic Metals, Proceedings of
  International Conference on Synthetic Metals, 12--18 July 1998,
  Montpellier, France} \\

\bigskip

  {\Large Temperature dependence of the normal-state Hall coefficient
  of a quasi-one-dimensional metal}

\bigskip

Victor M. Yakovenko and Anatoley T. Zheleznyak

\bigskip
Department of Physics and Center for Superconductivity Research,
University of Maryland, College Park, MD 20742, USA\\

\smallskip
\end{center}

{\normalsize\bf Abstract}

   \hspace*{4mm} We develop a systematic theory of the Hall effect in
Q1D conductors in both weak and strong magnetic fields for a model
where the electron relaxation time varies over the Fermi surface.  At
high temperatures, the Hall coefficient saturates at the value
$-\beta/ecn$, where the dimensionless coefficient $\beta$ is
determined by the curvature of the longitudinal dispersion law of
electrons, $e$ the electron charge, $c$ is the speed of light, and $n$
is the hole concentration.  At low temperatures, where a strong
variation of the relaxation rate over the Fermi surface develops in
the form of ``hot spots'', the Hall coefficient becomes
temperature-dependent and may change sign for a particular choice of
the transverse dispersion law parameters.  In our model, the sign
changes in a weak, but not in a strong magnetic field.


\bigskip\medskip
]

Recently, the $({\bf a},{\bf b})$-plane Hall resistivity was measured
under pressure in the normal state of (TMTSF)$_2$PF$_6$ and was found
to be strongly temperature-dependent and changing sign around 20 K
\cite{JeromeMoser-Private}.  We investigate a possible theoretical
explanation of this effect within the ``hot spots'' model
\cite{Yakovenko95a}, where the electron relaxation time strongly
varies over the Fermi surface.  The often-used formula for the Hall
coefficient $R_H=1/enc$ \cite{Abrikosov} (where $e$ and $n$ are the
electric charge and concentration of carriers, and $c$ is the speed of
light) applies only to the case of closed electron orbits in a strong
magnetic field and does not apply to quasi-one-dimensional (Q1D)
conductors, where the electron orbits are open.  In this paper, we
develop a systematic theory of the Hall effect in the latter case.

The (TMTSF)$_2$X materials can be modeled as layers of conducting
chains parallel to the $x$ axis and spaced at a distance $b$ along the
$y$ direction. A magnetic field $H$ is applied perpendicular to the
layers.  In this paper, we neglect coupling between the layers and
study a two-dimensional (2D) problem.  The results can be easily
generalized to 3D by integrating over the momentum component
perpendicular to the layers.

We start from the linearized stationary Boltzmann equation
\cite{Abrikosov} written for the deviation $\chi({\bf p})$ of the
electron distribution function $f({\bf p})$ from the equilibrium Fermi
function $f_0(\varepsilon)$: $f({\bf p})-f_0(\varepsilon)= -\chi({\bf
p})\,\partial f_0(\varepsilon)/\partial\varepsilon$, where {\bf p} and
$\varepsilon({\bf p})$ are the 2D electron momentum and energy:
\begin{equation}
  \frac{eH}{c}v(p_t)\frac{\partial\chi(p_t)}{\partial p_t}
  + \frac{\chi(p_t)}{\tau(p_t)} = e{\bf E} \cdot {\bf v}(p_t).
\label{eq:chi}
\end{equation}
In Eq.\ (\ref{eq:chi}), $p_t$ is the component of the electron
momentum tangential to the Fermi surface, which labels different
points on the Fermi surface.  ${\bf v}({\bf
p})=\partial\varepsilon({\bf p})/\partial{\bf p}$ and $\tau(p_t)$ are
the local values of the electron velocity and relaxation time at each
point $p_t$ on the Fermi surface.  {\bf E} is an applied electric
field, and $e$ is the (negative) electron charge.

In the case of a weak magnetic field $H$, the first term in Eq.\
(\ref{eq:chi}) is a small perturbation, so a solution can be obtained
as a series in powers of $H$ by iterating this term.  The Hall
conductivity per layer, $\sigma_{xy}$, is given by Ong's formula
\cite{Ong91}:
\begin{equation}
  \sigma_{xy} = -\frac{2e^3 H}{(2\pi\hbar)^2 c} \oint dp_t\,
  v_y(p_t)\tau(p_t)\,\frac{d[v_x(p_t)\tau(p_t)]}{dp_t},
\label{eq:sigma_xy}
\end{equation}
where the integral over $p_t$ is taken around the Fermi surface, and
the coefficient 2 comes from the two spin orientations.

We apply the general formula (\ref{eq:sigma_xy}) to a Q1D conductor,
where the Fermi surface consists of two open sheets located in the
vicinities of the the two longitudinal (along the chains) Fermi
momenta $\pm p_F$.  The electron dispersion laws in the vicinities of
$\pm p_F$ can be written as
\begin{eqnarray}
&& \varepsilon({\bf p})=\varepsilon_x(p_x) + \varepsilon_y(p_y),
\label{E_x+E_y} \\
&& \varepsilon_x=\pm v_F(p_x\mp p_F) - (p_x\mp p_F)^2/2m,
\label{E_x} \\
&& \varepsilon_y=2t_b\cos(k_y\pm\varphi) + 2t_b'\cos(2k_y\pm\varphi'),
\label{E_y}
\end{eqnarray}
where the energy $\varepsilon$ is measured from the Fermi energy, and
$k_y=p_yb/\hbar$.  The longitudinal energy dispersion law (\ref{E_x})
is expanded to the second power of $p_x\mp p_F$.  The coefficient of
the first term is the Fermi velocity $v_F$, and the coefficient of the
second term is
\begin{equation}
  1/m = -\partial^2\varepsilon_x/\partial p_x^2\,|_{p_F} = \beta\,v_F/p_F.
\label{eq:1/m}
\end{equation}
The longitudinal energy band in (TMTSF)$_2$X can be modeled as a
tight-binding band 1/4-filled by holes or as a parabolic band for
holes.  In these cases, the dimensionless coefficient $\beta$ in Eq.\
(\ref{eq:1/m}) has the following values:
\begin{equation}
  \beta=\left\{ \begin{array}{ll}
    \pi/4=0.785 & \quad {\rm for} \quad
      \varepsilon_x=2t_a\cos\left(p_xa/\hbar\right), \\
    1 & \quad {\rm for} \quad \varepsilon_x=-p_x^2/2m,
  \end{array} \right.
\label{eq:beta}
\end{equation}
where $t_a$ and $a$ are the electron tunneling amplitude and spacing
between the TMTSF molecules along the chains.  In Eq.\ (\ref{E_y}),
$t_b$ and $t_b'$ are the electron tunneling amplitudes between the
nearest and next-nearest chains, and the phases $\varphi$ and
$\varphi'$ originate from the triclinic structure of the (TMTSF)$_2$X
crystals \cite{Yamaji86}. (We otherwise ignore the triclinic
structure, considering the $x$, $y$, and $z$ axis to be mutually
orthogonal.)  Below, in Eqs.\ (\ref{eq:v_x})--(\ref{eq:sigma_yy}), we
calculate the transport contribution from the $+p_F$ Fermi sheet and
double the result to account for the $-p_F$ sheet.

A simple approximation where the electron relaxation time
$\tau(p_t)=\tau$ and longitudinal velocity $v_x(p_t)=v_F$ are assumed
to be constant over the Fermi surface is insufficient to calculate the
Hall conductivity, because Eq.\ (\ref{eq:sigma_xy}) gives
$\sigma_{xy}=0$ in this case.  Thus, we need to take into account the
second term $v_x^{(2)}$ in the expression for the longitudinal
velocity obtained from Eq.\ (\ref{E_x}) at the Fermi surface
$\varepsilon(p_x,p_y)=0$:
\begin{equation}
  v_x=v_F+v_x^{(2)},\quad v_x^{(2)}=-(p_x-p_F)/m\approx\varepsilon_y/mv_F.
\label{eq:v_x}
\end{equation}
Substituting the first term, $v_F$, into Eq.\ (\ref{eq:sigma_xy}),
integrating by parts, and averaging over $k_y$ instead of $p_t$:
\begin{equation}
\int_0^{2\pi}\frac{dk_y}{2\pi}\,f(k_y)=\langle f(k_y)\rangle_{k_y},
\label{eq:<>}
\end{equation}
we find one contribution to the Hall conductivity:
\begin{eqnarray}
  \sigma_{xy}^{(1)} &\;=\;& \frac{e^3Hv_F}{\pi\hbar^2c}\left\langle
  \tau^2(k_y)\,\frac{\partial v_y(k_y)}{\partial k_y}\right\rangle_{k_y}
\nonumber \\
  &=& \frac{e^3Hv_Fb}{\pi\hbar^3c}\left\langle
  \tau^2(k_y)\,\frac{\partial^2 \varepsilon_y(k_y)}{\partial k_y^2}
  \right\rangle_{k_y}.
\label{eq:sigma_xy_1}
\end{eqnarray}
Another contribution is obtained by substituting $v_x^{(2)}$ from Eq.\
(\ref{eq:v_x}) into Eq.\ (\ref{eq:sigma_xy}):
\begin{eqnarray}
  \sigma_{xy}^{(2)} &\;=\;& -\frac{2e^3 H}{\pi\hbar c bmv_F} 
  \left\langle\tau^2(k_y)\,v_y^2(k_y)\right\rangle_{k_y}
\nonumber \\
  &=& -\beta\frac{2e^3 Hb}{\pi\hbar^3cp_F} 
  \left\langle\tau^2(k_y)\left(
  \frac{\partial\varepsilon_y(k_y)}{\partial k_y}
  \right)^2\right\rangle_{k_y},
\label{eq:sigma_xy_2} \\
  \sigma_{xy}^{(3)} &=& \beta\frac{e^3 Hb}{2\pi\hbar^3cp_F} 
  \left\langle\tau^2(k_y)\,
  \frac{\partial^2\varepsilon_y^2(k_y)}{\partial k_y^2}\right\rangle_{k_y}.
\label{eq:sigma_xy_3}
\end{eqnarray}
Eq.\ (\ref{eq:sigma_xy_2}) originates from the derivative $\partial
v_x^{(2)}(k_y)/\partial k_y$ in Eq.\ (\ref{eq:sigma_xy}), and Eq.\
(\ref{eq:sigma_xy_3}) comes from the derivative
$\partial\tau(k_y)/\partial k_y$ integrated by parts.

For calculation of the diagonal components of the conductivity tensor,
it is sufficient to use $v_F$ for $v$ and $v_x$:
\begin{eqnarray}
  \sigma_{xx} &\;=\;& \frac{2e^2v_F}{\pi\hbar b}
   \left\langle\tau(k_y)\right\rangle_{k_y},
\label{eq:sigma_xx} \\
  \sigma_{yy} &=& \frac{2e^2}{\pi\hbar b v_F}
   \left\langle\tau(k_y)\,v_y^2(k_y)\right\rangle_{k_y}
\nonumber \\
  &=& \frac{2e^2b}{\pi\hbar^3 v_F}
  \left\langle\tau(k_y)\left(
  \frac{\partial\varepsilon_y(k_y)}{\partial k_y}
  \right)^2\right\rangle_{k_y}.
\label{eq:sigma_yy} 
\end{eqnarray}

Using Eqs.\ (\ref{eq:sigma_xy_1})--(\ref{eq:sigma_yy}), we calculate
the Hall coefficient:
\begin{equation}
  R_H =\sum_{i=1}^3 R_H^{(i)}, \quad {\rm where} \quad
  R_H^{(i)} = \frac{\sigma_{xy}^{(i)}} {H \sigma_{xx} \sigma_{yy}}.
\label{eq:R_H}
\end{equation}
Of the three terms $R_H^{(i)}$
(\ref{eq:sigma_xy_1})--(\ref{eq:sigma_xy_3}), only $R_H^{(2)}$ has a
nonzero value $R_H^{(0)}$ when $\tau(k_y)=\rm const$:
\begin{equation}
  R_H^{(2)}\to R_H^{(0)}=-\beta/ecn,
\label{eq:R_H_2}
\end{equation}
where $n=4p_F/2\pi\hbar b$ is the concentration of holes.  The minus
sign in Eq.\ (\ref{eq:R_H_2}) corresponds to the hole sign of the Hall
effect.  Eq.\ (\ref{eq:R_H_2}) differs from the conventional formula
$R_H=1/ecn$ by the coefficient $\beta$ given by Eq.\ (\ref{eq:beta}),
which is proportional to the curvature (nonlinearity) of the
longitudinal dispersion law (see Eq.\ (\ref{eq:1/m})).  Eq.\
(\ref{eq:R_H_2}) was obtained earlier in Ref.\ \cite{Cooper}.  For a
parabolic band, $\beta=1$, and Eq.\ (\ref{eq:R_H_2}) coincides with
the naive result.  However, for a 1/4-filled tight-binding model,
$R_H^{(0)}$ is reduced by the factor $\pi/4$.  The curvature $\beta$
and the Hall coefficient $R_H^{(0)}$ may be further reduced, if
dimerization of the TMTSF stack is taken into account.  For a
half-filled tight-binding model, $\beta$ and $R_H^{(0)}$ vanish.

\begin{figure}
\centerline{\psfig{file=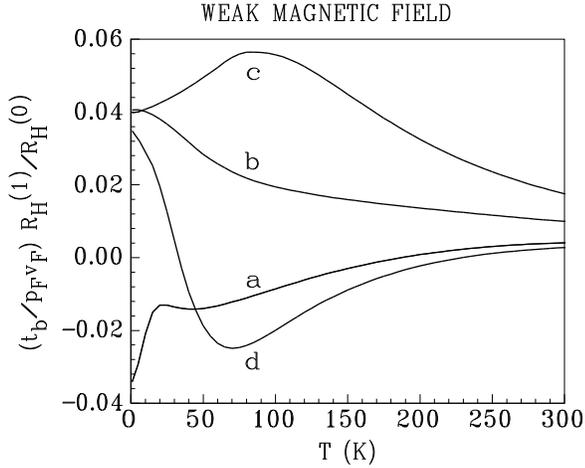,width=\linewidth,angle=-90}}
\caption{Temperature dependence of the first contribution $R_H^{(1)}$
  to the Hall coefficient, normalized to $(v_Fp_F/t_b)R_H^{(0)}$, in the
  case $\omega_c\tau\ll1$.}
\label{fig:R_H1w}
\end{figure}

\begin{figure}
\centerline{\psfig{file=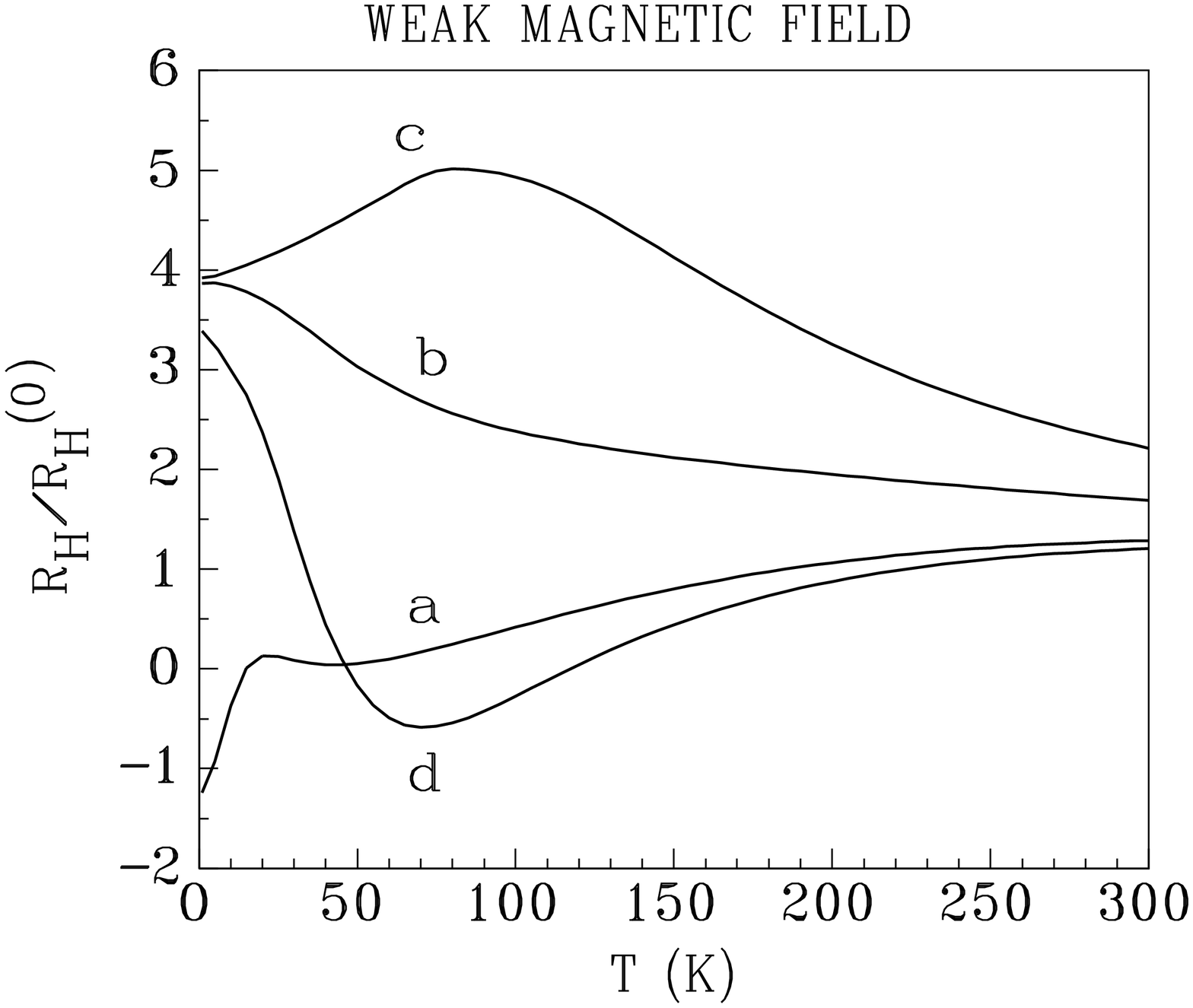,width=\linewidth,angle=-90}}
\caption{Temperature dependence of the Hall coefficient $R_H$,
  normalized to $R_H^{(0)}$, in the case $\omega_c\tau\ll1$.}
\label{fig:R_Hw}
\end{figure}

\begin{figure}
\centerline{\psfig{file=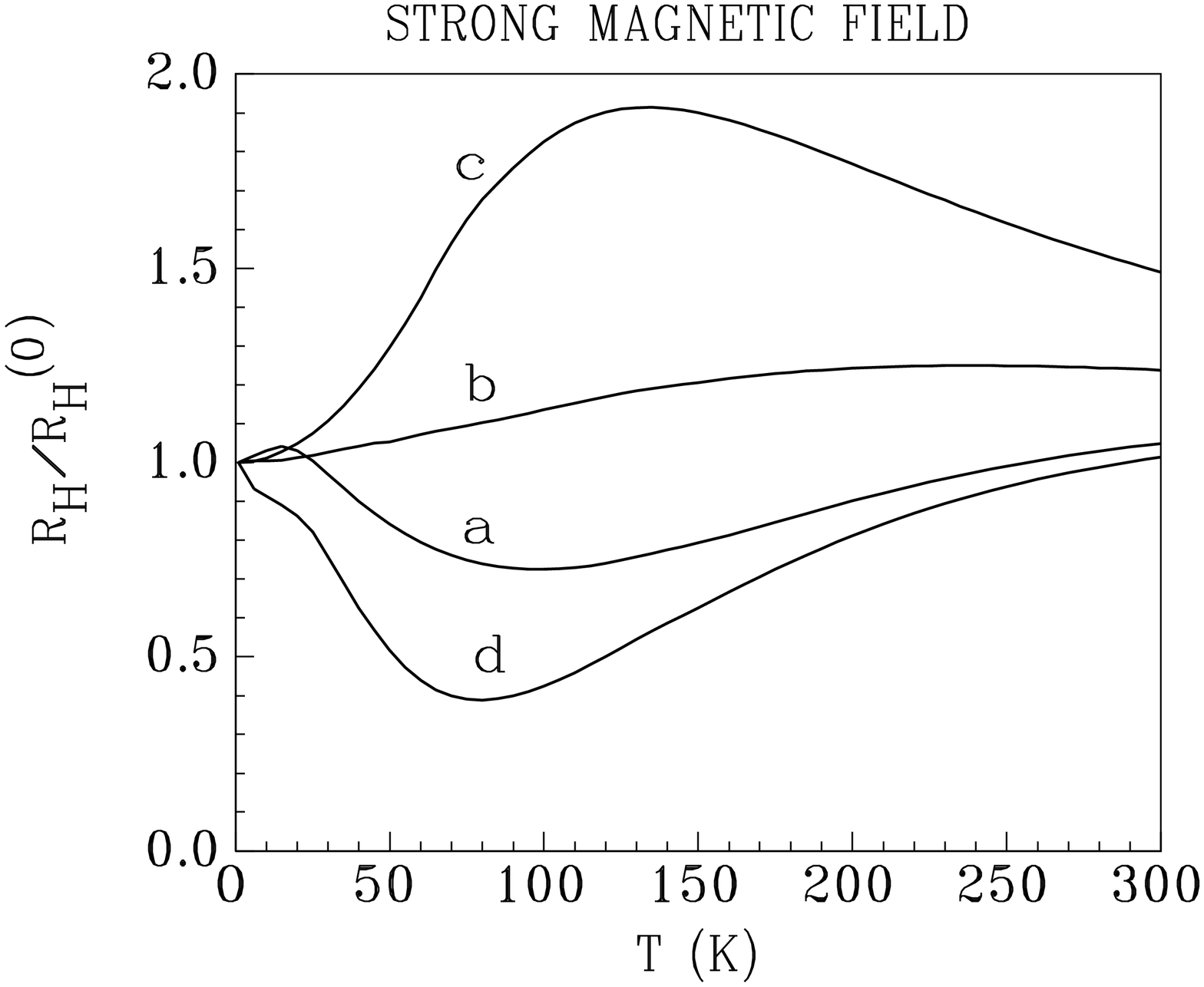,width=\linewidth,angle=-90}}
\caption{Temperature dependence of the Hall coefficient $R_H$,
  normalized to $R_H^{(0)}$, in the case $\omega_c\tau\gg1$.}
\label{fig:R_Hs}
\end{figure}

On the other hand, $R_H^{(1)}$ and $R_H^{(3)}$ are completely
determined by variation of the relaxation time $\tau(k_y)$ over the
Fermi surface.  $R_H^{(1)}$ also vanishes by symmetry when $t_b'=0$
because of $\tau(k_y)=\tau(k_y+\pi)$.  In Ref.\ \cite{Yakovenko95a},
we calculated the distribution of the electron scattering rate
$1/\tau(k_y)$ over the Fermi surface of a Q1D metal.  We found that a
strong variation of $1/\tau(k_y)$ develops at low temperatures, where
the relaxation rate at certain ``hot spots'' becomes much higher than
on the rest of the Fermi surface.  In the present paper, following the
same method, we compute the distribution of the electron relaxation
rate $1/\tau(k_y)$ for the dispersion law (\ref{E_x+E_y})--(\ref{E_y})
with $t_b=300$ K, $t_b'=30$ K, and several choices of $\varphi$ and
$\varphi'$.  In Figs.\ \ref{fig:R_H1w}--\ref{fig:R_Hs}, the curves (a)
correspond to $\varphi=\varphi'=0$, the curves (b) to
$\varphi=\varphi'/2=\pi/8$, the curves (c) to
$\varphi=\varphi'/2=\pi/4$, and the curves (d) to $\varphi=\pi/12$ and
$\varphi'=0$.

We substitute the computed $\tau(k_y)$ into Eqs.\
(\ref{eq:sigma_xy_1})--(\ref{eq:R_H}) and calculate $R_H^{(i)}$ as a
function of temperature $T$.  On one hand, $\sigma_{xy}^{(1)}$
(\ref{eq:sigma_xy_1}) is relatively small, because it requires an
asymmetry in the $\tau(k_y)$ distribution.  On the other hand, it is
proportional to the first power of the transverse tunneling amplitude
$t_b$, whereas $\sigma_{xy}^{(2)}$ (\ref{eq:sigma_xy_2}) is
proportional to $t_b^2$.  Thus, $\sigma_{xy}^{(1)}$ is enhanced
relative to $\sigma_{xy}^{(2)}$ by the big factor $v_Fp_F/t_b$ equal
to $\pi t_a/\sqrt{2}t_b$ in the 1/4 filled tight-binding model of
(TMTSF)$_2$X.  In Fig.\ (\ref{fig:R_H1w}), we show the temperature
dependence of $R_H^{(1)}$ normalized to $(v_Fp_F/t_b)R_H^{(0)}$ (with
$\beta=\pi/4$).  We observe that $R_H^{(1)}$ is strongly
temperature-dependent at low temperatures $T\leq t_b=300$ K because of
development of the ``hot spots'' in $\tau(k_y)$ and vanishes at high
temperatures $T\geq t_b$.  The different curves in Fig.\
\ref{fig:R_H1w} illustrate the sensitivity of $R_H^{(1)}$ to the
choice of the phases $\varphi$ and $\varphi'$ in the transverse
dispersion law (\ref{E_y}), which determines the pattern of
$\tau(k_y)$.

The temperature dependence of the total Hall coefficient $R_H(T)$,
normalized to $R_H^{(0)}$, is shown in Fig.\ \ref{fig:R_Hw} for the
1/4-filled band with $t_a/t_b=30$, which corresponds to
$p_Fv_F/t_b=66.6$.  We observe that all curves saturate at $R_H^{(0)}$
at high temperatures, and the curves (a) and (d) change sign at lower
temperatures.  The total Hall coefficient $R_H(T)$ may change sign
only for a sufficiently high anisotropy $t_a/t_b$, which makes
$R_H^{(1)}$ comparable with $R_H^{(0)}$.

The above results apply in the limit of a weak magnetic field where
$\omega_c\tau\ll1$.  The cyclotron frequency $\omega_c=2\pi/t_0$ is
defined in terms of the time $t_0$ it takes an electron to circle
around the Fermi surface in the case of a closed orbit or traverse the
Brillouin zone in the case of an open orbit.  However, the magnetic
fields 4--12 T used in the experiment \cite{JeromeMoser-Private} most
likely correspond to the limit of a strong magnetic field:
$\omega_c\tau\gg1$.  We study this case below.

A general analytical solution of Eq.\ (\ref{eq:chi}) gives the
following expression for the conductivity tensor \cite{Abrikosov}:
\begin{eqnarray}
&& \sigma_{ij}=\frac{2ec}{(2\pi\hbar)^2H}\,
   \frac{1}{1-\exp[-(c/eH)\oint dp_t/v(p_t)\tau(p_t)]}
\label{eq:sigma_ij} \\
&& \times \oint dp_t\,\frac{v_i(p_t)}{v(p_t)} 
  \oint\limits_{p_t-P_t}^{p_t} dp_t'\,\frac{v_j(p_t')}{v(p_t')}
   \exp\left(\frac{-c}{eH}\int\limits_{p_t'}^{p_t}
   \frac{dp_t''}{v(p_t'')\tau(p_t'')} \right),
\nonumber
\end{eqnarray}
where $P_t$ is the period of integration over the tangential momentum
$p_t$.  Expanding Eq.\ (\ref{eq:sigma_ij}) to the first power of
$1/H$, we find the following expression for the Hall conductivity in
the limit of a strong magnetic field:
\begin{eqnarray}
  \sigma_{xy}&\;=\;&-\frac{2ec}{(2\pi\hbar)^2H\oint dp_t/v(p_t)\tau(p_t)}
\nonumber \\
  &&\times \oint dp_y \oint\limits_{p_t-P_t}^{p_t} dp_x'\,
  \int\limits_{p_t'}^{p_t}\frac{dp_t''}{v(p_t'')\tau(p_t'')},
\label{eq:Sigma_xy_general}
\end{eqnarray}
where $dp_x$ and $dp_y$ are the $x$ and $y$ projections of the
differential $dp_t$.  Taking the integral over $dp_x'$ by parts, we
find:
\begin{eqnarray}
  \sigma_{xy} &\;=\;& \frac{2ec}{(2\pi\hbar)^2H}
\label{eq:Sigma_xy_II} \\
  && \times\left(\oint dp_y\,p_x
  -\frac{1}{\oint dp_t/v\tau} \oint dp_y 
  \oint\limits_{p_t-P_t}^{p_t}\frac{dp_t'\,p_x'}{v(p_t')\tau(p_t')}
  \right).
\nonumber
\end{eqnarray}
For closed orbits, the integral $\oint dp_y$ in the second term in
Eq.\ (\ref{eq:Sigma_xy_II}) vanishes, and the first term produces the
familiar formula $\sigma_{xy}=enc/H$ \cite{Abrikosov}.  However, the
situation is the opposite for open orbits in Q1D conductors.  In this
case, the first term in Eq.\ (\ref{eq:Sigma_xy_II}) vanishes, if $p_x$
is counted from the Fermi momentum $p_F$, whereas the second term does
not vanish and gives the following expression:
\begin{equation}
  \sigma_{xy}=-\frac{2ec}{\pi\hbar Hb}\,
  \frac{1}{\langle1/v_x(k_y)\tau(k_y)\rangle_{k_y}}\left\langle
  \frac{p_x(k_y)-p_F}{v_x(k_y)\tau(k_y)}\right\rangle_{k_y}.
\label{eq:Sigma_xy}
\end{equation}
In this and the following equations
(\ref{eq:Sigma_xy})--(\ref{eq:Sigma_xx}), (\ref{eq:Sigma_yy}), and
(\ref{eq:Sigma_yy'}), the averaging over $k_y$ is performed for the
$+p_F$ sheet of the Fermi surface, and the result is doubled to
account for the $-p_F$ sheet.

The Hall conductivity (\ref{eq:Sigma_xy}) vanishes if $\tau(k_y)=\tau$
and $v_x(k_y)=v_F$ are assumed to be constant over the Fermi surface.
Thus, we need to take into account both terms in Eq.\ (\ref{eq:v_x}),
which produce the following two contributions to the Hall conductivity
when substituted into Eq.\ (\ref{eq:Sigma_xy}):
\begin{equation}
  \sigma_{xy}^{(1)}=\frac{2ec}{\pi\hbar Hv_Fb}\,
  \frac{1}{\langle1/\tau(k_y)\rangle_{k_y}}\left\langle
    \frac{\varepsilon_y(k_y)}{\tau(k_y)}\right\rangle_{k_y},
\label{eq:Sigma_xy_1}
\end{equation}
\begin{equation}
  \sigma_{xy}^{(2)}=-\beta\frac{2ec}{\pi\hbar Hp_Fv_F^2b}\,
  \frac{1}{\langle1/\tau(k_y)\rangle_{k_y}}\left\langle
    \frac{\varepsilon_y^2(k_y)}{\tau(k_y)}\right\rangle_{k_y}.
\label{eq:Sigma_xy_2}
\end{equation}

The longitudinal conductivity is given by the zeroth-order term of
expansion of Eq.\ (\ref{eq:Sigma_xy_II}) in powers of $1/H$:
\begin{equation}
  \sigma_{xx}
  =\frac{2e^2}{\pi\hbar b\langle1/v_x(k_y)\tau(k_y)\rangle_{k_y}}
  \approx\frac{2e^2v_F}{\pi\hbar b\langle1/\tau(k_y)\rangle_{k_y}}.
\label{eq:Sigma_xx}
\end{equation}
The transverse conductivity is given by the second-order term of
expansion of Eq.\ (\ref{eq:Sigma_xy_II}) in powers of $1/H$.  The term
originating from expansion of the first line in Eq.\ 
(\ref{eq:sigma_ij}) is similar to Eq.\ (\ref{eq:Sigma_xy_II}) with the
integral over $p_y$ replaced by an integral over $p_x$.  This term
vanishes. Another term, originating from expansion of the second line
in Eq.\ (\ref{eq:sigma_ij}), has the following form:
\begin{eqnarray}
  \sigma_{yy}&\;=\;&
  \frac{c^2}{(2\pi\hbar)^2H^2\oint dp_t/v(p_t)\tau(p_t)}
\nonumber \\
  &&\times\oint dp_x \oint\limits_{p_t-P_t}^{p_t} dp_x' \left(
  \int\limits_{p_t'}^{p_t}\frac{dp_t''}{v(p_t'')\tau(p_t'')}
  \right)^2.
\label{eq:Sigma_yy_general}
\end{eqnarray}
Taking the integrals over $p_x'$ and $p_x$ by parts, we find:
\begin{eqnarray}
  \sigma_{yy} &\;=\;&
  \frac{2c^2}{\pi\hbar H^2b\langle1/v_x(k_y)\tau(k_y)\rangle_{k_y}} 
\label{eq:Sigma_yy} \\
  &&\times \left(\left\langle\frac{(p_x-p_F)^2}{v_x\tau}\right\rangle_{k_y}
  \left\langle\frac{1}{v_x\tau}\right\rangle_{k_y}
  -\left\langle\frac{p_x-p_F}{v_x\tau}\right\rangle_{k_y}^2 \right).
\nonumber
\end{eqnarray}
The second term in the brackets in Eq.\ (\ref{eq:Sigma_yy}) vanishes
when $\tau(k_y)=\rm const$ and $v_x(k_y)=\rm const$, whereas the first
term does not.  Thus, the second term is much smaller that the first
term and can be neglected.  Besides, the second term exactly cancels
with $\sigma_{xy}^2$ (\ref{eq:Sigma_xy}) in the combination
$\sigma_{xx}\sigma_{yy}+\sigma_{xy}^2$ that appears in
$R_H=\sigma_{xy}/H(\sigma_{xx}\sigma_{yy}+\sigma_{xy}^2)$.  Thus, in
Q1D conductors with open electron orbits,
$\sigma_{xy}^2\ll\sigma_{xx}\sigma_{yy}$ even in the limit
$\omega_c\tau\gg1$, unlike in metals with closed electron orbits.
Neglecting the second term in Eq.\ (\ref{eq:Sigma_yy}), we find:
\begin{equation}
  \sigma_{yy}\approx\frac{2c^2}{\pi\hbar H^2v_F^3b} 
  \left\langle\frac{\varepsilon_y^2(k_y)}{\tau(k_y)}\right\rangle_{k_y}.
\label{eq:Sigma_yy'}
\end{equation}

Using Eqs.\ (\ref{eq:Sigma_xy_1}), (\ref{eq:Sigma_xy_2}),
(\ref{eq:Sigma_xx}), and (\ref{eq:Sigma_yy'}), we find two
contributions to the Hall coefficient:
\begin{equation}
  R_H = R_H^{(1)} + R_H^{(2)} = 
  \frac{\sigma_{xy}^{(1)}}{H\sigma_{xx}\sigma_{yy}}
  + \frac{\sigma_{xy}^{(2)}}{H\sigma_{xx}\sigma_{yy}}.
\label{eq:R_H'}
\end{equation}
In Fig.\ \ref{fig:R_Hs}, we show the temperature dependence $R_H(T)$
in the case $\omega_c\tau\gg1$ calculated for the same model with
$t_a/t_b=30$ as in Fig.\ \ref{fig:R_Hw}.  At high temperatures, where
$\tau(k_y)$ does not depend on $k_y$, $R_H^{(1)}$ vanishes, whereas
$R_H^{(2)}$ saturates at $R_H^{(0)}$ given by Eq.\ (\ref{eq:R_H_2}).
At lower temperatures, where the ``hot spots'' in $\tau(k_y)$ develop,
$R_H^{(1)}$ and, consequently, $R_H$ become strongly
temperature-dependent.  However, $R_H^{(1)}$ vanishes at $T=0$, and
$R_H$ returns to the value $R_H^{(0)}$.  This is related to the
peculiar expression for the relaxation rate due to umklapp scattering
along the chains at $T=0$ in our model \cite{Yakovenko95a}:
\begin{eqnarray}
1/\tau(k_y)&\;\propto\;&T^2\int dk_y^{(1)}\,dk_y^{(2)}\,dk_y^{(3)}\,
\delta(k_y+k_y^{(1)}-k_y^{(2)}-k_y^{(3)})
\nonumber \\
&&\times\delta[\varepsilon_y^+(k_y)+\varepsilon_y^+(k_y^{(1)})
  +\varepsilon_y^-(k_y^{(2)})+\varepsilon_y^-(k_y^{(3)})],
\label{eq:1/tau}
\end{eqnarray}
where $\varepsilon_y^+(k_y)$ and $\varepsilon_y^-(k_y)$ are the
transverse dispersion laws (\ref{E_y}) for the $\pm p_F$ sheets.  The
average over $k_y$ that appears in Eq.\ (\ref{eq:Sigma_xy_1})
vanishes, because it can be written as an average of the argument of
the delta-function:
\begin{eqnarray}
\langle\varepsilon_y&&(k_y)/\tau(k_y)\rangle_{k_y}
\propto(1/4)\int dk_y\,dk_y^{(1)}\,dk_y^{(2)}\,dk_y^{(3)}\,
\nonumber \\
&&\times\delta(k_y+k_y^{(1)}-k_y^{(2)}-k_y^{(3)})
\nonumber \\
&&\times[\varepsilon_y^+(k_y)+\varepsilon_y^+(k_y^{(1)})
  +\varepsilon_y^-(k_y^{(2)})+\varepsilon_y^-(k_y^{(3)})]
\nonumber \\
&&\times\delta[\varepsilon_y^+(k_y)+\varepsilon_y^+(k_y^{(1)})
  +\varepsilon_y^-(k_y^{(2)})+\varepsilon_y^-(k_y^{(3)})]=0.
\label{eq:0}
\end{eqnarray}

In conclusion, we have developed a systematic theory of the Hall
effect in Q1D conductors for the model where the electron relaxation
time $\tau(k_y)$ varies over the Fermi surface. We have studied the
cases of both weak ($\omega_c\tau\ll1$) and strong
($\omega_c\tau\gg1$) magnetic fields.  At high temperatures, the Hall
coefficient saturates at the value $-\beta/ecn$, where the
dimensionless coefficient $\beta$ is determined by the second
derivative $\partial^2\varepsilon_x/\partial p_x^2$ of the
longitudinal dispersion law of electrons (see Eqs.\ (\ref{eq:1/m}) and
(\ref{eq:beta})).  At low temperatures, where a strong variation of
the relaxation rate over the Fermi surface develops in the form of
``hot spots'', $R_H$ becomes temperature-dependent and may change sign
for a particular choice of the transverse dispersion law parameters.
In our model, the sign changes in a weak, but not in a strong magnetic
field.

We need to add two caveats to our study.  First, the electron
relaxation rate in this paper and in Ref.\ \cite{Yakovenko95a} was
calculated for umklapp scattering along the chains.  While this
scattering is believed to be relevant for calculating $\sigma_{xx}$,
it is not clear whether is is also relevant for calculating of
$\sigma_{xy}$ and $\sigma_{yy}$.  On the other hand, our Eq.\
(\ref{eq:chi}) requires to use the same relaxation time for all
transport coefficients.  Because of this uncertainty, our numerical
calculations of $R_H$ should be considered only as a qualitative
illustration.  Second, the sign change of $R_H(T)$ was experimentally
found in Ref.\ \cite{JeromeMoser-Private} in the regime where the
temperature dependence of resistivity changes slope from metallic to
insulating, which may be due to opening of a charge pseudogap
\cite{Jerome95}.  Our conventional, Fermi-liquid treatment of
transport coefficients may not apply in this complicated situation.

\vspace{-1\baselineskip}

\end{document}